\documentstyle[12pt,epsfig,psfig]{article}
\newcommand {\ignore}[1]{}

\newcommand{\Frac}[2]{\frac{\displaystyle #1}{\displaystyle #2}}

\newcommand{\noi}{\noindent}
\newcommand{\bc}{\begin{center}}
\newcommand{\ec}{\end{center}}

\def\ifmath#1{\relax\ifmmode #1\else $#1$\fi}

\def\3quarter{{\textstyle{3 \over 4}}}

\def\ra{\rightarrow}

\def\lf{\leaders\hbox to 1em{\hss.\hss}\hfill}
\def\21{$SU(2) \ot U(1)$}
\def\321{$SU(3) \ot SU(2) \ot U(1)$}

\def\nt{\hbox{$\nu_\tau$ }}

\def\mnt{\hbox{$m_{\nu_\tau}$ }}

\def\ie{\hbox{\it i.e., }}

\def\gau{\hbox{gauge }}

\def\eq#1{{eq. (\ref{#1})}}

\def\fig#1{{Fig. (\ref{#1})}}

\def\VEV#1{\left\langle #1\right\rangle}

\def\ltap{\raisebox{-.4ex}{\rlap{$\sim$}} \raisebox{.4ex}{$<$}}
\def\gtap{\raisebox{-.4ex}{\rlap{$\sim$}} \raisebox{.4ex}{$>$}}
\def\lsim{\raise0.3ex\hbox{$\;<$\kern-0.75em\raise-1.1ex\hbox{$\sim\;$}}}
\def\gsim{\raise0.3ex\hbox{$\;>$\kern-0.75em\raise-1.1ex\hbox{$\sim\;$}}}

\def\beq{\begin{equation}}
\def\eeq{\end{equation}}
\def\bef{\begin{figure}}
\def\eef{\end{figure}}
\def\bet{\begin{table}}
\def\eet{\end{table}}
\def\bea{\begin{eqnarray}}
\def\ba{\begin{array}}
\def\ea{\end{array}}
\def\bi{\begin{itemize}}
\def\ei{\end{itemize}}
\def\ben{\begin{enumerate}}
\def\een{\end{enumerate}}
\def\ra{\rightarrow}
\def\ot{\otimes}
\def\eea{\end{eqnarray}}
\def\nps#1#2#3{        {\it Nucl. Phys. B (Proc. Suppl.) }{\bf #1} (19#2) #3} 
\def\np#1#2#3{           {\it Nucl. Phys. }{\bf #1} (19#2) #3}
\def\pl#1#2#3{           {\it Phys. Lett. }{\bf #1} (19#2) #3}
\def\pr#1#2#3{           {\it Phys. Rev. }{\bf #1} (19#2) #3}
\def\prep#1#2#3{         {\it Phys. Rep. }{\bf #1} (19#2) #3}
\def\prl#1#2#3{          {\it Phys. Rev. Lett. }{\bf #1} (19#2) #3}

\def\cpc#1#2#3{         {\it Comp. Phys. Commun. }{\bf #1} (19#2) #3}

\def\Sci#1#2#3{          {\it Science }{\bf #1} (19#2) #3}
\def\n.c.#1#2#3{         {\it Nuovo Cim. }{\bf #1} (19#2) #3}
\def\r.n.c.#1#2#3{       {\it Riv. del Nuovo Cim. }{\bf #1} (19#2) #3}

\def\ppnp#1#2#3{           {\it Prog. Part. Nucl. Phys. }{\bf #1} (19#2) #3}

\def\ptmm{$p\!\!\!/_T +\mu^+ \mu^-$}

\def\cpc#1#2#3{         {\it Comp. Phys. Commun. }{\bf #1} (19#2) #3}

\begin{document}
\thispagestyle{empty}
\begin{titlepage}
\begin{center}
{\Large \bf LEP Sensitivities to Spontaneous 
R-Parity Violating Signals}
\vskip 0.2cm
{\Large J. C. Rom\~ao}
\footnote{E-mail fromao@alfa.ist.utl.pt}\\
{\it Inst. Superior Tecnico, Dept. de F\'{\i}sica \\
Av. Rovisco Pais, 1 - 1096 Lisboa, Codex, PORTUGAL}\\
\vskip 0.2cm
{\Large F. de Campos} \footnote{E-mail fernando@axp.ift.unesp.br}\\
{\it Instituto de F\'{\i}sica Te\'orica, \\
Universidade Est. Paulista, \\
Rua Pamplona, 145, 01405-900, S\~ao Paulo, BRAZIL}\\
\vskip 0.2cm
{\Large M. A. Garc\'{\i}a-Jare\~no}
\footnote{E-mail miguel@flamenco.ific.uv.es}, 
{\Large M. B. Magro}\footnote{E-mail magro@flamenco.ific.uv.es} and 
{\Large J. W. F. Valle}
\footnote{E-mail valle@flamenco.ific.uv.es}\\
{\it Instituto de F\'{\i}sica Corpuscular - C.S.I.C.\\
Departament de F\'{\i}sica Te\`orica, Universitat de Val\`encia\\
46100 Burjassot, Val\`encia, SPAIN\\
URL: http://neutrinos.uv.es}\\
\vskip .2cm
{\bf Abstract}
\end{center}
\begin{quotation}
\noindent

We illustrate the sensitivities of LEP experiments to leptonic 
signals associated to models where supersymmetry (SUSY) is 
realized with spontaneous breaking of R-parity. We focus on 
missing transverse momentum plus acoplanar muon events
($ p\!\!\!/_T +\mu^+ \mu^- $) arising from lightest 
neutralino single production $\chi \nu$ as well as pair
production $\chi \chi$, followed by $\chi$ decays, where 
$\chi$ denotes the lightest neutralino.  We show that the 
integrated luminosity achieved at LEP already starts probing the 
basic parameters of the theory. We discuss the significance of 
these constraints for the simplest spontaneous R-parity breaking 
models and their relevance for future searches of SUSY particles.

\end{quotation}
\end{titlepage}
\section{Introduction}
So far most searches for supersymmetric particles have been made 
in the framework of the {\sl Minimal Supersymmetric Standard Model 
(MSSM)} which assumes the conservation of a discrete symmetry called 
R-parity \cite{mssm}. Under this symmetry all the standard model 
particles are R-even while their superpartners are R-odd. R-parity 
is related to the spin (S), total lepton number (L), and baryon number 
(B) according to $R_p=(-1)^{(3B+L+2S)}$. In the limit of exact R-parity 
the supersymmetric (SUSY) particles must be produced only in pairs, 
the lightest of them being stable. 

Unfortunately there is no clear dynamical clue as to how supersymmetry
is realized. In fact, neither gauge invariance nor supersymmetry
require the conservation of R-parity. The violation of R-parity could 
emerge as the residual effect of a more fundamental unified theory
\cite{expl0}. While R-parity violation may be explicit \cite{expl},
we find it rather attractive to consider the possibility that it 
arise in a spontaneous way, like the breaking of the electroweak 
symmetry \cite{aul}.
At the present state-of-the-art theory can not decide. It is therefore 
of great interest to pursue the phenomenological implications of 
alternative scenarios. This is specially so in view of the fact that 
the associated phenomena can be accessible to experimental verification.

Recently there was a lot of attention devoted to the possibility 
that R-parity can be an exact symmetry of the Lagrangian, broken 
spontaneously through nonzero vacuum expectation values (VEVS) for 
scalar neutrinos \cite{MASIpot3,MASI,ROMA,ZR,RPCHI,mono,NPBTAU,RPMSW}.  
There are two main types of scenario \cite{beyond}.
If lepton number is part of the gauge symmetry there is an 
additional gauge boson which acquires mass via the Higgs mechanism.
In this case there is no physical Goldstone boson and the scale of 
R-parity violation also characterises the new gauge interaction, 
around the TeV scale \cite{ZR}. Consequently, its effects can be 
large \cite{RPCHI}. In this model typically the lightest SUSY 
particle (LSP) is a neutralino which decays mostly to visible states,
breaking R-parity. The main decay modes are three-body,
\beq
\label{vis}
\chi \ra f \bar{f} \nu
\eeq
where f denotes a charged fermion. Its invisible decay modes
are in the channel 
\beq
\label{invi}
\chi \ra 3 \nu
\eeq
Alternatively, if spontaneous R-parity violation occurs 
in the absence of any additional gauge symmetry, it leads to the 
existence of a physical massless Nambu-Goldstone boson, called
Majoron (J) \cite{fae}. Thus in this case {\sl the lightest SUSY 
particle is the Majoron} which is massless and therefore stable
\footnote{The majoron may have a small mass due to explicit
breaking effects at the Planck scale. In this case it may decay 
to neutrinos and photons. However the time scales are only of 
cosmological interest and do not change the signal expected at 
the laboratory \cite{KEV}.}. As a result often the lightest 
neutralino $\chi$ may decay invisibly, conserving R-parity, as 
\beq
\label{invis}
\chi \ra \nu + J.
\eeq
With the minimal particle content of the MSSM, including only the 
usual isodoublet sneutrinos, these models lead to a new decay mode 
for the $Z$ boson, $Z \ra \rho \: + \: J$, where $\rho$ is a light 
scalar with mass $\ll M_W$. This decay increases the invisible width 
of the $Z$ by the equivalent of one half extra neutrino species, 
and is therefore ruled out by the LEP measurements. 
This difficulty, as well as the fine-tuning problem characteristic
of this simplest scenario, is naturally avoided by adding \21 singlet 
neutrinos, in such a way so that the R-parity breaking is driven by the 
corresponding sneutrino VEVS. In this case the Majoron is mainly singlet 
\cite{MASIpot3,MASI}.

In this paper we analyse some aspects of the phenomenology of 
spontaneously broken R-parity models at the Z peak. 
We consider the production of the lightest supersymmetric 
fermion, including both single as well as pair production mechanisms. 
In order to identify the corresponding signals we take
into account all possible neutralino decay channels, thus 
extending previous discussions.
We focus on the study of events with missing transverse momentum 
plus acoplanar muons ($ p\!\!\!/_T +\mu^+ \mu^- $), arising both
from neutralino single production at the Z peak
\beq
Z \ra \chi \nu \:  \:, \:  \: 
\eeq
as well as pair production mechanisms
\beq
Z \ra \chi \chi \:,
\eeq
followed by the R-parity violating $\chi$ decays in \eq{vis} or 
\eq{invi} and \eq{invis}.
Using the integrated luminosities attained by the four LEP experiments 
we conclude that the basic parameters of the model are starting to be probed 
in a meaningful way. We discuss the theoretical significance of these 
constraints from the point of view of other R-parity-violating processes
as well as their possible implications for future SUSY particle searches.

\section{Basic Structure}

Most of our subsequent analysis will be very general and applies
to a wide class of \21 models with spontaneously broken R-parity, 
such as those of ref. \cite{MASIpot3,MASI}, as well as models 
where the majoron is absent due to an enlarged gauge structure 
\cite{ZR,RPCHI}. Many of the phenomenological features relevant 
for the LEP studies discussed here already emerge in an effective 
model where the violation of R-parity is introduced explicitly 
through a bilinear superpotential term of the type $\ell H_u$ 
\cite{epsi}.

For concreteness, we start by adopting the conceptually
simplest model for the spontaneous violation of R-parity 
proposed in ref. \cite{MASIpot3} and by recalling its basic 
ingredients. This will serve to set up our notation in what follows. 
We consider this model as the most useful 
way to parametrise the resulting physics, due to the strict 
correlation exhibited in this model between the magnitude of 
R-parity violating phenomena and the \nt mass. For example,
all single SUSY particle production rates as well as the lightest 
neutralino decay rate $\Gamma_\chi$ are directly correlated to the 
mass of the tau neutrino. 
The superpotential is given by
\beq
h_u  Q H_u u^c + h_d  H_d Q d^c + h_e \ell H_d e^c +
(h_0 H_u H_d - \epsilon^2) \Phi +
h_{\nu} \ell H_u \nu^c  + h \Phi  S \nu^c + h.c.
\label{P}
\eeq
This superpotential conserves $total$ lepton number and R-parity.
The superfields $(\Phi$, ${\nu^c}_i$, $S_i)$ are singlets under
\21 and carry a conserved lepton number assigned as
$(0,-1,1)$ respectively. The couplings $h_u,h_d,h_e,h_{\nu},h$ are
arbitrary matrices in generation space. The additional singlets 
$\nu^c, S$ \cite{SST} and $\Phi$ \cite{BFS} may drive the spontaneous 
violation of R-parity. This leads to the existence 
of a Majoron given by the imaginary part of \cite{MASIpot3}
\beq
\frac{v_L^2}{Vv^2} (v_u H_u - v_d H_d) +
              \frac{v_L}{V} \tilde{\nu_{\tau}} -
              \frac{v_R}{V} \tilde{\nu^c}_{\tau} +
              \frac{v_S}{V} \tilde{S_{\tau}}
\label{maj}
\eeq
where the isosinglet VEVS
\beq
\begin{array}{lr}
v_R = \VEV {\tilde{\nu}_{R\tau}}\:, &
v_S = \VEV {\tilde{S_{\tau}}}
\end{array} 
\eeq
with $V = \sqrt{v_R^2 + v_S^2}$ characterise R-parity or lepton 
number breaking and the isodoublet VEVS
\beq
\begin{array}{lr}
v_u = \VEV {H_u} \:, &
v_d = \VEV {H_d} 
\end{array} 
\eeq
are responsible for the breaking of the electroweak symmetry and
the generation of fermion masses. The combination $v^2 = v_u^2 + v_d^2$ 
is fixed by the $W,Z$ masses. Finally, there is a small seed of R-parity 
breaking in the doublet sector, \ie
\beq
v_L = \VEV {\tilde{\nu}_{L\tau}} 
\eeq
whose magnitude is now related to the Yukawa coupling
$h_{\nu}$. Since this vanishes as $h_{\nu} \ra 0$, 
we can naturally obey the limits from stellar energy 
loss \cite{KIM}. 

For our subsequent discussion we need the chargino and neutralino 
mass matrices. The form of the chargino mass matrix is common to 
a wide class of \21 SUSY models with spontaneously broken R-parity
and is given by
\beq
\begin{array}{c|cccccccc}
& e^+_j & \tilde{H^+_u} & -i \tilde{W^+}\\
\hline
e_i & h_{e ij} v_d & - h_{\nu ij} v_{Rj} & \sqrt{2} g_2 v_{Li} \\
\tilde{H^-_d} & - h_{e ij} v_{Li} & \mu & \sqrt{2} g_2 v_d\\
-i \tilde{W^-} & 0 & \sqrt{2} g_2 v_u & M_2
\end{array}
\label{chino}
\eeq
Two matrices U and V are needed to diagonalise the $5 \times 5$ 
(non-symmetric) chargino mass matrix
\bea
{\chi}_i^+ = V_{ij} {\psi}_j^+\\
{\chi}_i^- = U_{ij} {\psi}_j^-
\label{INO}
\eea
where the indices $i$ and $j$ run from $1$ to $5$ and
$\psi_j^+ = (e_1^+, e_2^+ , e_3^+ ,\tilde{H^+_u}, -i \tilde{W^+}$)
and $\psi_j^- = (e_1^-, e_2^- , e_3^-, \tilde{H^-_d}, -i \tilde{W^-}$).

Under reasonable approximations, we can truncate the neutralino 
mass matrix so as to obtain an effective $7\times 7$ matrix of 
the following form \cite{MASIpot3}
\beq
\begin{array}{c|cccccccc}
& {\nu}_i & \tilde{H}_u & \tilde{H}_d & -i \tilde{W}_3 & -i \tilde{B}\\
\hline
{\nu}_i & 0 & h_{\nu ij} v_{Rj} & 0 & g_2 v_{Li} & -g_1 v_{Li}\\
\tilde{H}_u & h_{\nu ij} v_{Rj} & 0 & - \mu & -g_2 v_u & g_1 v_u\\
\tilde{H}_d & 0 & - \mu & 0 & g_2 v_d & -g_1 v_d\\
-i \tilde{W}_3 & g_2 v_{Li} & -g_2 v_u & g_2 v_d & M_2 & 0\\
-i \tilde{B} & -g_1 v_{Li} & g_1 v_u & -g_1 v_d & 0 & M_1
\end{array}
\label{nino}
\eeq
This matrix is diagonalised by a $7 \times 7$ unitary matrix N,
\beq
{\chi}_i^0 = N_{ij} {\psi}_j^0
\eeq
where 
$\psi_j^0 = ({\nu}_i,\tilde{H}_u,\tilde{H}_d,-i \tilde{W}_3,-i \tilde{B}$),
with $\nu_i$ denoting the three weak-eigenstate neutrinos.

In the above two equations $M_{1,2}$ denote the supersymmetry 
breaking gaugino mass parameters and $g_{1,2}$ are the 
$SU(2) \ot U(1)$ \gau couplings divided by $\sqrt{2}$.
We assume the canonical relation $M_1/M_2 = \frac{5}{3} tan^2{\theta_W}$.
Note that the effective Higgsino mixing parameter $\mu$ may be 
given in some models as $\mu = h_0 \VEV \Phi$, where $\VEV \Phi$ 
is the VEV of an appropriate singlet scalar.

Typical values for the SUSY parameters $\mu$, $M_2$, 
and the parameters $h_{\nu i,3}$ lie in the range given by 
\beq
\label{param2}
\begin{array}{cccc}
-250\leq\frac{\displaystyle\mu}{\mbox{GeV}}\leq 250 & & & 
30 \leq \frac{\displaystyle M_2}{\mbox{GeV}}\leq 1000 \\
&&& \\
10^{-10}\leq h_{\nu 13},h_{\nu 23} \leq 10^{-1} & & & 
10^{-5}\leq h_{\nu 33} \leq 10^{-1}\\
\end{array}
\eeq
while the expectation values lie in the range:
\beq
\label{param1}
\begin{array}{cccc}
v_L=v_{L3}=100\:\: \mbox{MeV} & & & 
v_{L1}=v_{L2}=0 \\
&&& \\
50\:\: \mbox{GeV}\leq v_R=v_{R3}\leq 1000 \:\: \mbox{GeV} \ & & & 
v_{R1}=v_{R2}=0\\
&&& \\
50\:\: \mbox{GeV}\leq v_S=v_{S3}=v_R\leq 1000 \:\: \mbox{GeV} 
& & & 1 \leq \tan\beta=\frac{v_u}{v_d} \leq 30  \\
\end{array}
\eeq 
The diagonalization of \eq{nino} gives rise to the mixing of the 
neutralinos with the neutrinos, leading to R-parity violating
gauge couplings. In what follows we will give explicit 
expressions for the couplings of the SUSY fermions in 
terms of these diagonalizing matrices.

\section{Charged and Neutral Current Couplings}

Using the above diagonalizing matrices U, V and N one can write the 
electroweak currents of the mass-eigenstate fermions.
For example, the charged current Lagrangian describing the 
weak interaction between charged lepton/chargino and
neutrino/neutra\-linos may be written as
\beq
\frac{g}{\sqrt2} W_\mu \bar{\chi}_i^- \gamma^\mu 
(K_{Lik} P_L + K_{Rik} P_R) {\chi}_k^0 + H.C.
\label{CC}
\eeq
where $P_{L,R}$ are the two chiral projectors and
the $5\times 7$ coupling matrices $K_{L,R}$ may
be written as
\bea
K_{Lik} = \eta_i (-\sqrt2 U_{i5} N_{k6} - U_{i4}
N_{k5} - \sum_{m=1}^{3} U_{im}N_{km})\label{KL}\\
K_{Rik} = \epsilon_k (-\sqrt2 V_{i5} N_{k6} + V_{i4} N_{k4})
\label{KR}
\eea
The matrix $K_{Lik}$ is the analogous of the matrix $K$ introduced 
in ref \cite{2227} and there is a corresponding matrix characterizing
right-handed charged currents $K_{Rik}$. These "off-diagonal" blocks
corresponding to $i=1..3$ and $k=4..7$ as well as $i=4,5$ and $k=1..3$
are R-parity-breaking couplings.

The corresponding neutral current Lagrangian may be written as
\beq
\frac{g}{\cos\theta_W} Z_\mu \{ \bar{\chi}_i^- \gamma^\mu 
(O'_{Lik} P_L + O'_{Rik} P_R) \chi_k^- 
+ \frac{1}{2} 
\bar{\chi}_i^0 \gamma^\mu 
( O''_{Lik} P_L + O''_{Rik} P_R) \chi_k^0 
\}
\label{NC}
\eeq
where the $7 \times7$ coupling matrices $O'_{L,R}$ and
$O''_{L,R}$ are given by
\bea
O'_{Lik} =\eta_i \eta_k \left( \frac{1}{2} U_{i4} U_{k4} + U_{i5}
U_{k5} + \frac{1}{2} \sum_{m=1}^{3} U_{im}U_{km} -\delta_{ik}
\sin^2\theta_W \right)\label{OL}\\
O'_{Rik} = \frac{1}{2} V_{i4} V_{k4} + V_{i5}
V_{k5} - \delta_{ik} \sin^2\theta_W \label{OR}\\
O''_{Lik} =\frac{1}{2} \epsilon_i \epsilon_k   \left( N_{i4} N_{k4} - N_{i5}
N_{k5} - \sum_{m=1}^{3} N_{im}N_{km} \right) = - \epsilon_i \epsilon_k
O''_{Rik} \label{O11R}
\eea
The off-diagonal part of these coupling matrices break R-parity,
i.e. when $i=4..7$ and $k=1..3$ or vice-versa.

In writing these  couplings we have assumed CP conservation. 
Under this assumption the diagonalizing matrices can be chosen 
to be real. The $\eta_i$ and $\epsilon_k$ factors are sign 
factors, related with the relative CP parities of these 
fermions, that follow from the diagonalization of their 
mass matrices. 

Like all supersymmetric extensions of the standard model,
the spontaneously broken R-parity models are constrained by 
data that follow from the negative searches for supersymmetric
particles at LEP, in particular the most recent limits on 
chargino masses from the recent run at 130 GeV mass region
\cite{LEPSEARCH} as well as $\bar pp$ collider data gluino 
production \cite{D0}.

There are additional restrictions, which are 
more characteristic of broken R-parity models. They follow 
from laboratory experiments related to neutrino physics and 
weak interactions, cosmology and astrophysics \cite{fae,beyond}.
These restrictions play a very important role, as they exclude 
many parameter choices that are otherwise allowed by the collider 
constraints, while the converse is not true. The most relevant 
constraints come from neutrinoless double beta decay and neutrino 
oscillation searches, direct searches for anomalous peaks at $\pi$
and K meson decays, the limit on the tau neutrino mass \cite{eps95},
cosmological limits on the \nt lifetime and mass, as well as limits 
on muon and tau lifetimes, on lepton flavour violating decays
and universality violation.

One can perform a sampling of the points in our model which 
are allowed by all of the above constraints in order to evaluate 
systematically the attainable value of the couplings \cite{RPLHC}. 
The allowed values for the diagonal (R-parity conserving) 
couplings for the lightest neutralino $\chi$ and the lightest 
chargino $\chi^\pm$ are of the same order as those in the MSSM. 

\ignore{
The coupling of the lightest chargino to the $Z$ is maximum when 
it is mainly a gaugino. In this case $\mu \gg M_2$ and therefore 
$|V_{45}|\approx |U_{45}|\approx 1$ and $|V_{4i}|,|U_{4i}|\ll 1$ 
for $i\neq 5$. 
On the other hand it is minimum when its larger component is along the 
Higgsino. In this case $\mu \ll M_2$ and therefore 
$|V_{44}|\approx |U_{44}|\approx 1$ and 
$|V_{4i}|,|U_{4i}|\ll 1$ for $i\neq 4$. Including these values 
in \eq{OL} and \eq{OR} one gets for the allowed range
\beq
\label{zdc}
0.27\ltap |O'_{L44}|, |O'_{R44}|\ltap 0.77
\eeq
From the parameters given in \eq{param2} and \eq{param1}
and the experimental limits from LEP (specially the limit on the
lightest chargino mass) one finds that the lightest supersymmetric 
fermion is always a neutralino, with mass $ M_{\chi^0}  \gtap 25 \: GeV $. 

The R-parity-conserving $W$ coupling of the lightest $chargino$ 
to the lightest $neutralino$ is maximal when the chargino consists 
mainly of a Higgsino ($|V_{44}|\approx |U_{44}|\approx 1$ and
$|V_{4i}|,|U_{4i}|\ll 1$ for $i\neq 4$). In that case there are 
two light neutralinos which form a Quasi-Dirac state 
($|N_{44}|\approx |N_{45}|\approx \frac{1}{\sqrt{2}}$). Including 
these values of the diagonalizing matrices in \eq{KL} and \eq{KR} one gets the 
upper bound on these couplings.
Experimental constraints (see next section) determine the lower 
limit. With all that we have the allowed range
\beq
\label{wd}
10^{-4} \ltap |K_{L44}|,|K_{R44}| \ltap \frac{1}{\sqrt{2}}
\eeq
}

For the neutral current couplings of the lightest neutralino 
$\chi$ we can only get an upper limit
\ignore{This is because they vanish both 
in the limits $\mu \gg M_2$ and $\mu \ll M_2$. In the first case 
since the lightest neutralino is mainly a gaugino it does not couple 
to the $Z$. In the second case the lightest neutralino is 
mainly a Quasi-Dirac Higgsino and its couplings cancel almost 
exactly CHECK.}
After imposing the experimental constraints explained before, we get 
\beq
\label{zdn}
|O''_{L44}| \ltap  0.1
\eeq
In what concerns the R-parity breaking couplings, the biggest ones 
correspond to the standard lepton belonging to the third family, i.e. 
$O''_{L43}$, and this will be responsible for the $Z \ra \chi \nt$ 
decay. This coupling amplitude can reach a few per cent or so for
neutralino mass values accessible at LEP \cite{beyond} and, as
we will show, may lead decay rates observable at LEP. 

In what follows we will focus on some of the zen-event signals that 
could be associated to neutralino single as well as pair production 
and subsequent decays at LEP. 

\section{Neutralino Production at the Z Peak}

At the LEP I collider the neutralinos may
be produced as 
\beq
\label{prod}
 e^+ e^- \ra \chi_i \chi_j 
\eeq
The differential cross section for these processes including only
the $Z$ exchange at $\sqrt{s}=M_Z$ is given by
\begin{eqnarray}
\label{x}
\frac{d\sigma}{d\Omega} ( e^+ e^- \ra \chi_i \chi_j ) &=& 
\frac{\alpha^2}{4s} \frac{1}{2}(2-\delta_{i j})|Q(s)|^2 
\lambda^{1/2} \left(1, \frac{m_i^2}{s}, \frac{m_j^2}{s}\right) 
\left(\frac{1}{\sin{\theta_w} \cos{\theta_w}}\right)^4\nonumber\\
&& \left[ G_{1ij} (s)+ G_{2ij} (s) \cos \theta + G_{3ij} (s)
\cos^2\theta \right]
\end{eqnarray}
where $\lambda$ is the usual K\"allen function and
\begin{eqnarray}
\label{x1}
G_{1ij}(s) & = & \left(g^2_V + g^2_A \right)
\left[ 2 \frac{E_i}{\sqrt{s}} \frac{E_j}{\sqrt{s}}
\left(O^{''2}_{Lij} + O^{''2}_{Rij}\right) + 4 
\frac{m_i m_j}{s} O^{''}_{Lij} O^{''}_{Rij} \right] \\
G_{2ij}(s) &=& 2 g_V g_A \left(O^{''2}_{Lij} - O^{''2}_{Rij}\right)
\lambda^{1/2} \left(1, \frac{m_i^2}{s}, \frac{m_j^2}{s}\right) \\
G_{3ij} (s) &=& \frac{1}{2} \left( g^2_V + g^2_A \right)
            \left(O^{''2}_{Lij} + O^{''2}_{Rij}\right) 
\lambda \left(1, \frac{m_i^2}{s}, \frac{m_j^2}{s}\right)
\end{eqnarray}
and
\begin{eqnarray}
\label{x2}
&&E_i = \frac{s + m_i^2 - m_j^2}{2\sqrt{s}} \,\,;\,\, 
E_j = \frac{s + m_j^2 - m_i^2}{2\sqrt{s}} \\
&&Q(s) = \frac{s}{s-M_Z^2+iM_Z\Gamma_Z}
\end{eqnarray}
Here $g_V$ and $g_A$ are the usual vector and axial couplings for 
the Standard Model $Ze^+e^-$ vertex 
\begin{equation}
g_V=-\frac{1}{4} + \sin^2 \theta_w \qquad ; \qquad
g_A=-\frac{1}{4}
\end{equation}
and the relevant $\chi_i$ $\chi_j$ 
coupling amplitudes are determined from \eq{O11R}. Notice that, due 
to the Majorana nature of the neutralinos, their neutral current couplings 
obey the relation $O^{''}_{Rij}=-\epsilon_i \epsilon_j
O^{''}_{Lij}$ and therefore the functions $G_{2ij}(s)$ vanish identically.
In the above equations the indices $i$ and $j$ run from $1$ to $7$. 
At $\sqrt{s}=M_{Z}$ they will be restricted only to those 
corresponding to neutralinos lighter than the $Z$ boson. 

As seen above, in models with spontaneously broken R-parity,
the mixing of the standard leptons with the supersymmetric 
charginos and neutralinos leads to the existence of R-violating 
couplings in the Lagrangian when written in terms of the mass eigenstates. 
As a result, SUSY particles can be singly-produced. 
This means that $\chi_i$ and $\chi_j$ in \eq{prod} can be both 
supersymmetric particles (standard SUSY pair production) as well as 
one standard and one supersymmetric (R-parity breaking single production). 
This is in sharp contrast with explicitly broken R-parity models 
such as considered in ref. \cite{aleph95}, where only the pair production 
of the lightest neutralino $\chi$ can take place at the Z peak.

In this paper we are therefore concerned with the signals arising from 
$Z \ra \chi \nu$ decays (corresponding to $i=j=4$) and 
$Z \ra \chi \chi$ decays (corresponding to $i=4, j=3$ or vice-versa), 
where $\chi$ denotes the lightest neutralino. The heavier ones will be 
assumed to be too heavy to be produced at the Z peak.

\section{Neutralino Decays}

Once produced in $e^+ e^-$ collisions, the neutralino subsequently 
decays, typically inside the detector. In order to identify the expected 
signatures at LEP it is necessary to specify its possible decay modes. 
In the MSSM all supersymmetric particles have cascade decays 
finishing in the LSP which is normally a neutralino. However,
if R-parity is broken there are new decay channels and the 
supersymmetric particles can decay directly to the standard 
states breaking R-parity. Also the lightest SUSY particle
may not be a neutralino, in fact this is the case in 
spontaneously broken models \cite{MASIpot3,MASI}
\footnote{
In this case the LSP may be produced in R-parity-violating decays
of normal particles like the muon, the tau lepton \cite{NPBTAU}
or the Z boson \cite{mono}.}. 
Alternatively, SUSY particles may decay 
through R-parity conserving cascade decays that will produce
the lightest neutralino $\chi$. This may decay invisibly conserving
R-parity as in \eq{invis}, or via the R-parity breaking three 
fermion modes in \eq{vis}. To the extent that the invisible 
dominates, as could happen, one recovers the lightest neutralino 
missing momentum signal expected in the MSSM.

For simplicity we are going to study the decays of the lightest 
neutralino, which one expects would be the earliest-produced 
supersymmetric particle. Heavier states would have cascade 
decays that we are not going to consider here. 

The lightest neutralino $\chi$ has to decay always to standard states breaking 
R-parity. If its mass is lower than the mass of the gauge bosons it decays 
to the three body final states. 
\ignore{
\beq
\begin{array}{lllll}
\chi^0\ra \nu_j f \bar f & &  \mbox{with width} & & 
\Gamma^0_{13bj}=8(v_f^2+a_f^2)
\Gamma^{3b}(M_{\chi^0},0,M_Z,O''_{L4j},O''_{R4j})\\
\chi^0\ra l_j f_u \overline{f_d} & &  \mbox{with width} & & 
\Gamma^0_{23bj}=\Gamma^{3b}(M_{\chi^0},0,M_W,K_{Lj4},K_{Rj4})\\
\end{array}
\eeq
For the decay $\chi^0\ra \nu_j  l_j^+ l_j^-$ there is interference between 
the charged and neutral current and the width is given by
\beq 
\Gamma^0_{33bj}=\Gamma^{3b\prime}(M_{\chi^0},K_{Lj4},K_{Rj4},
O''_{L4j},O''_{R4j})
\eeq
The explicit expressions for the widths $\Gamma^{3b}$ and
$\Gamma^{3b\prime}$ are given in the Appendix, and correct some
errors in ref. \cite{RPLHC}. 
}
We will distinguish three cases. First all the particles in the final
state are neutral. Then
\beq
\chi^0 \ra \nu_j\ \nu_k\ \nu_m  \hskip 1cm
\mbox{with width} \hskip 1cm 
\Gamma^0_{1}=\Gamma^{3b}(M_{\chi^0},O''_{L},O''_{R})
\label{3neutrinos}
\eeq
In the second case we consider charged leptons in the final state,
that is
\beq
\chi^0\ra \nu_j\ l^-_k\ l^+_m \hskip 0.5cm
\mbox{with width} \hskip 0.5cm 
\Gamma^0_{2}=\Gamma^{3b'} 
(M_{\chi^0},O''_{L},O''_{R},O'_{L},O'_{R},K_{L},K_{R})
\label{3leptons}
\eeq
In both these processes there is interference among the various
diagrams contributing. This is clear for the process in
Eq.~\ref{3leptons} where there is interference between the neutral and
charged current diagrams, but it also true for the process in
Eq.~\ref{3neutrinos} due to the Majorana nature of the neutrinos. 
The explicit expressions for the widths $\Gamma^{3b}$ and
$\Gamma^{3b\prime}$ are given in the Appendix.
Finally there is a third case when the final state contains quarks,
that is
\beq
\chi^0\ra \nu_j\ q \ \overline{q} \hskip 0.5cm
\mbox{with width} \hskip 0.5cm 
\Gamma^0_{3}=\Gamma^{3b'} 
(M_{\chi^0},O''_{L},O''_{R},O'_{L},O'_{R},0,0)
\eeq
that proceeds only via neutral current. In the Appendix it is
explained how this width can be obtained as a particular case of
$\Gamma^{3b\prime}$.

As mentioned in the introduction, the existence of the Majoron implies 
that in \21 spontaneously broken R-parity, the neutralino can always decay 
invisibly \eq{invis} with a decay width 
\beq
\label{wJ}
\begin{array}{l}
\Gamma^0_{Jj}=\frac{1}{32\pi}M_{\chi^0} (C_{L4j}^2+C_{R4j}^2) \\
C_{Lij}=- \epsilon_i \epsilon_j C_{Rij}={\displaystyle\sum_{k=1}^3}
\epsilon_j (N_{ik} N_{j4}+  N_{jk} N_{i4}) h_{\nu k3} 
\Frac{v_R}{\sqrt{2}V} \\
\end{array}
\eeq
Although our discussion will be more general, we neglect,
for definiteness, supersymmetric fermion decays mediated 
by slepton exchange. In this approximation, neutralinos
of mass accessible at LEP have only three-body decay modes 
mediated by charged and neutral currents, except for the 
two-body majoron decay \eq{invis}, characteristic of the simplest 
spontaneous R-parity breaking models. 

\section{Signals at LEP}

In order to study the experimental signals associated to the
first kinematically accessible neutral supersymmetric fermions, 
we have developed an event generator that simulates the 
processes expected for the LEP collider at $\sqrt{s}=M_{Z}$. 
It allows us to estimate the detection efficiencies when 
suitable selection criteria are imposed in order to avoid 
the expected standard model backgrounds to the processes 
of interest. We describe below the main steps we follow 
in order to generate neutralino production and decays.
As far as the production is concerned, our generator 
simulates the following processes at the $Z$ peak:
\begin{itemize}
\item $a)\ e^+ e^-\rightarrow \chi\nu$
\item $b)\ e^+ e^-\rightarrow \chi \chi$ 
\end{itemize}

Process a) clearly violates R-parity, so it is a new mode of
neutralino production forbidden in the MSSM, as well as in models
of explicitly broken R-parity \cite{expl} other than those where
this violation is due to a bilinear superpotential term $\ell H_u$ 
\cite{epsi}. On the other hand, process b) is allowed both in the 
MSSM as well as in models such as the one used in ref. \cite{expl}. 

The second step of the generation is the decay of the lightest neutralino,
which is the most characteristic feature of the R-parity breaking models. 
As explained in the introduction, if the lightest neutralino $\chi$ is
lighter than the $Z$ boson it will have three body decays via charged 
or neutral currents, as well as the two body invisible 
decay into neutrino + majoron.
These decays produce new supersymmetric signals and the generator 
allows their detailed study. In contrast, the two-body neutralino
decay into neutrino + majoron has as signal missing transverse 
momentum, because both final particles escape detection and therefore
we do not need to generate this process. 
Thus it suffices for us to generate the three body 
neutralino decays:
\begin{itemize}
\item 
$\chi \rightarrow \nu_{\tau} Z^{*} \rightarrow \nu_{\tau}\ l^+l^- ,
\nu_{\tau} \nu \nu,\nu_{\tau} q_i \overline{q_i}$  
\item 
$\chi \rightarrow \tau \ W^{*} \rightarrow  \tau \nu_i l_i,\tau q_u
\overline{q_d}$ 
\end{itemize}
The last step of our simulation is made calling the 
PYTHIA software \cite{pythia}, using as input 
the neutralino decay products above mentioned.

The signals associated to the first kinematically accessible 
neutralino, which arise from its single 
R-parity violating production as well as its R-parity conserving 
pair production are listed in table 1. This table shows the final 
signals for the $\chi \nu_{\tau}$ and $\chi\chi$
production with the subsequent $\chi$ decay.
One of the cleanest and most interesting signals that can be 
studied are the events with missing transverse momentum + 
acoplanar muons ($ p\!\!\!/_T +\mu^+ \mu^- $). These can be 
produced through either process a) or  
process b) as shown in table 1.

The main source of 
background for this signal is the $Z$ decay to $\mu^+\mu^-$ 
with the radiation of a $\gamma$ which may escape detection. 
This background has to be eliminated through suitable cuts.
For definiteness we have imposed the cuts used by the OPAL 
experiment for their search for acoplanar dilepton events 
\cite{opal91}:
\begin{itemize}
\item 
we select events with two muons with at least for one of the muons
obeying $ |\cos\theta | $ less than 0.7. 
\item 
the energy of each muon has to be greater than a $6\%$ of the beam
energy.
\item 
the missing transverse momentum in the event must exceed
$6\%$ of the beam energy, $p\!\!\!/_T > 3 \:$ GeV.
\item 
the acoplanarity angle (the angle between the projected
momenta of the two muons in the plane orthogonal to the 
beam direction) must exceed $20^o$. 
\end{itemize}

The OPAL experiment did not find any acoplanar muon pair
event passing these cuts in the data sample analysed in ref.
\cite{opal91}.

Using our Monte Carlo generator we did the calculation of the
kinematic distributions before and after applying these cuts.
This was used to determine the detection efficiencies associated
to the $\chi\nu$  and $\chi\chi$. 
They lie in the range $\approx 25\% -18\%$ 
for neutralino masses in the range $m_{\chi} \approx 25 GeV - 40 GeV$ 
for the case of $\chi\chi$ production. For the single 
production case ($\chi\nu$) we have found efficiencies in the range 
$11 \% -5\%$ for $m_{\chi} \approx 40 GeV -80 GeV$.

\section{Analysis and Results}

Given a sample of LEP data collected at the $Z$ peak, the study of the 
signal described in the previous section allows us to determine the 
corresponding experimental limits on the values of the relevant R 
parity-violating couplings versus neutralino mass. In order to 
illustrate the procedure we will use the last data published by 
the ALEPH collaboration, corresponding to an integrated luminosity 
$\displaystyle{L_{int}={number\ hadronic\ events\over 
\sigma(e^+e^-\rightarrow hadrons)}}=82\ pb^{-1}$ \cite{aleph95}.
This in turn corresponds to $1.94 \times 10^6$ hadronic decays of the Z. 
Of course this is justified only under the assumption that the 
experimental cuts of the previous section are enough to eliminate 
all relevant background, or that the detection efficiency for 
our signal is not reduced by further cuts that might be needed. 
Clearly, in order to obtain rigorous limits for a given data 
sample collected in a given experiment, one would have to check
whether these assumptions are true by means of more detailed 
simulation studies of the corresponding detector features as well as the
corresponding background for the given luminosity. 

While we await for a more complete statistics to be analysed
\cite{new96} we find it useful to illustrate the sensitivity 
to the basic parameters of our R-parity violating models which
has already been achieved with the data samples collected.
For such illustrative purposes we make use only of the cleanest
leptonic signal and use the published ALEPH data given in ref. 
\cite{aleph95}.
As usual, in order to obtain a $ 95\% CL $ limit on
some parameter, we impose the condition 
\begin{equation}
\label{3}
3 > N_{expt}
\end{equation}
$N_{expt}$ being the number of expected events for our signal, 
when no events of the desired type have been observed.

For the single production process $e^+e^- \rightarrow \chi \nu$ 
with the decay $\chi \rightarrow \nu_\tau \mu^+\mu^-$ the expected 
number \ptmm events is given as
\begin{equation}
N_{expt} (\chi \nu)  = \sigma(e^+e^-\rightarrow \chi \nu) BR(\chi \rightarrow 
\nu_\tau \mu^+\mu^-) \epsilon_{\chi \nu} \  L_{int}
\end{equation}
where $\epsilon_{\chi\nu}$ is the detection efficiency, obtained from 
the R-parity breaking generator  described before.

Using the expression for the cross section in \eq{x} we can write
\begin{eqnarray}
\label{mm1}
N_{expt} (\chi \nu) &=& \frac{2}{3} O^{''2}_{L43} 
\frac{\alpha^2 \pi (g_V^2+g_A^2)}{\Gamma_Z^2 (sin\theta_w cos\theta_w)^4}
 (2 - 3x_Z^2 + x_Z^6) \nonumber\\
& & BR (\chi \rightarrow \nu_\tau \mu^+\mu^-) \epsilon_{\chi\nu}  
\  L_{int}
\end{eqnarray}
where $x_Z = m_\chi/m_Z$.

In addition, the relation between the coupling $O^{''2}_{L43}$ and the
$BR(Z\ra\chi\nu)$ is given by
\begin{eqnarray}
\label{mm2}
BR(Z \ra \chi \nu) &=& \frac{2}{3}O^{''2}_{L43} \frac{M_Z^3 G_F}
{\Gamma_Z \pi \sqrt{2}} \left( 1-\frac{3}{2}x_Z^2+\frac{1}{2}x_Z^6 \right)
\end{eqnarray}
From \eq{mm1}, \eq{mm2} and \eq{3} one can  obtain a 
$95\% CL $ limit on the R-parity breaking observable
$BR (Z \ra \chi \nu) BR(\chi \rightarrow \nu_\tau \mu^+ \mu^-)$ 
as a function of the $\chi$ mass. This is shown in figures 2 and 4.

Here we should stress that this production mode is characteristic of
models with spontaneous violation of R-parity \cite{MASIpot3,
MASI,RPCHI}, or models that parametrize it through
an effective bilinear superpotential term $\epsilon \ell H$
\cite{epsi}.  It is absent in most models of explicitly broken 
R-parity, such as the one considered in the analysis
presented by the ALEPH collaboration in ref. \cite{aleph95}. 

Notice that the R-parity-violating parameter $\epsilon$
is directly correlated to the mass of the tau neutrino \nt.
As a result it is correspondingly restricted by cosmological
Big Bang nucleosynthesis \cite{bbncrisis}. The corresponding
limits are rather stringent \cite{bbmnu} and may not allow
\mnt above a few hundred KeV. In this case the signal displayed
in Fig. 4 would never reach $10^{-8}$. This cosmological bound can 
be avoided in models which contain neutrino decay \cite{unstable} or 
annihilation channels \cite{DPRV} beyond those present within 
the standard model. 
For this reason the spontaneously broken R-parity
models are preferred, as they allow the maximum
\mnt values permitted by laboratory experiments \cite{eps95} 
due to the \nt decays and/or annihilations
to majorons. However, as seen in Fig. 4, the maximum value
of our signal is only a bit larger than  $10^{-7}$. 

We conclude that in models such as the ones in ref. \cite{epsi,RPCHI}
one can probe the spontaneous violation of R-parity in the 
single-production mode for $m_\chi \approx 40$ GeV. However,
as already mentioned, these values of the signal rate
are hard to reconcile with cosmological \nt limits.
The corresponding rates for a cosmologically
acceptable \nt mass lie well bellow the dotted curve in
\fig{III}, below $10^{-8}$ for all values of the neutralino mass,
thus too small a rate to be of interest. Including the majoron,
whose existence is expected in any model where the spontaneous
violation of R-parity is realized in the minimal \21 gauge 
structure \cite{MASIpot3,MASI}, changes the situation 
in two ways. First, it improves the allowed signal rates because 
it allows \nt masses as large as present laboratory limits.
 Unfortunately there is a counter-effect that decreases the 
expected signal rates, because the presence of the invisible 
channel \eq{invis} tends to dilute the $\chi$ branching ratio into muons. 
The net result leads to the dotted curve in \fig{III}. 

The lightest neutralino $\chi$ may however be light enough to be 
pair-produced in the standard R-parity-conserving process 
$e^+e^- \ra\chi \chi$. Even in this case of standard production,
the violation of R-parity can provide visible
signals from the subsequent decays of the neutralinos.
The first possibility to consider here is the case where both 
neutralinos decay visibly, e.g. into, $\nu_\tau \mu^+\mu^-$. We do not
consider this possibility as it is similar to the one used 
recently by ALEPH and the corresponding sensitivities may be 
estimated by re-scaling the results of ref. \cite{aleph95}. 
Thus we choose concentrate on the novel possibility
that one of the neutralinos decays to $\nu_\tau \mu^+\mu^-$ 
while the other decays invisibly, which is more
characteristic of models with spontaneous
violation of R-parity. The number of expected
$p\!\!\!/_T+\mu^+ \mu^-$  events in this case is given by
\begin{equation}
N_{expt} (\chi \chi) = \sigma (e^+e^- \rightarrow \chi \chi)
2 BR (\chi \rightarrow \mbox{invisible})
BR(\chi\rightarrow\nu_\tau\mu^+\mu^-) \epsilon_{\chi \chi} \  L_{int}
\end{equation}
so that from \eq{x} we obtain
\begin{eqnarray}
N_{expt} (\chi \chi) &=& \frac{2}{3}O^{''2}_{L44} {\alpha^2 \over  \Gamma_Z^2} 
\frac{\pi (g_V^2 + g_A^2)}{(sin\theta_w cos\theta_w)^4}
(1-4x_Z^2)^{3/2}  \nonumber\\
&&2BR(\chi \rightarrow \nu_\tau \mu^+\mu-) BR(\chi \ra \mbox{invisible})
\epsilon_{\chi \chi} \  L_{int} 
\end{eqnarray}
and the corresponding expression for the $Z \ra \chi \chi$
branching ratio is
\begin{eqnarray}
BR(Z\ra\chi \chi ) &=& \frac{1}{3}O^{''2}_{L44}\frac{M_Z^3 G_F}{\Gamma_Z 
\pi \sqrt{2}} \left( 1- 4x_Z^2 \right)^{3/2}
\end{eqnarray}

From these last expressions and  \eq{3} it is possible to obtain 
an illustrative $95\% CL$ limit on $BR (Z \ra \chi \chi) 
BR (\chi \ra \nu_\tau \mu^+\mu^-) BR (\chi \ra \mbox{invisible})$ 
as a function of the $\chi$ mass, as shown in figures 3 and 5. 

\section{Discussion}

Using the integrated luminosity corresponding to the last published 
ALEPH data, $L= 82 \:pb^{-1}$, we have illustrated how the existing 
data gathered by the LEP collaborations at the Z peak are sufficient 
to start probing in a theoretically meaningful way the mass and 
couplings of the lowest lying neutral supersymmetric
fermions in spontaneously broken R-parity models. 
We have determined the corresponding regions of sensitivity, 
illustrated in fig. 2-5, both for the case of single production 
as well as for the case where the lightest neutralino can be pair 
produced. The theoretical significance of these decays is
illustrated in \fig{V}. 

From curve b in this \fig{III} one sees that the expected signal rate
for a neutralino of 40 GeV in the model of ref. \cite{MASIpot3} 
could be tested if the luminosities of the four LEP experiments
are added, even if only leptonic $p\!\!\!/_T +\mu^+ \mu^- $
and $p\!\!\!/_T +e^+ e^- $ channels are considered. This 
constraint would be significant from the point of view of the
underlying model. Indeed, as can be seen from \fig{V}, this
corresponds to values of the \nt mass close to an MeV, for which
other processes such as $Z \ra \chi^\pm + \tau^\mp$
\cite{ROMA,RPMSW} would be expected to be sizeable.
Although the presence of the majoron leads to novel decays
of standard model particles, e.g. $\mu \ra e + J$ and $\tau \ell + J$ 
decays \cite{NPBTAU,RPMSW} of relevance for intense muon source studies at 
PSI or for a future tau-charm factory \cite{tcf}, as well as novel Z decays 
such as $Z \ra \gamma + J$ \cite{mono}, its overall effect insofar as 
the signal rate for \ptmm events arising from $\chi\nu$ production
at the Z peak is concerned is to decrease. One advantage is that
the rate there is no conflict between the signal rate (dotted
curve in Fig. 3) and the primordial Helium abundance.
However, there is clearly a wide range of parameters still
unconstrained by this single production mechanism. In contrast,
if kinematically accessible, neutralino pair-production can place 
more severe restrictions, especially on models where
the spontaneous violation of R-parity where the majoron is
absent \cite{ZR} or in an effective model where the
violation of R-parity is introduced explicitly via the bilinear
superpotential term $\ell H_u$ \cite{epsi}. The constraints
derived in this case should be important in relation to
searches at higher energies such as LEP200, the NLC or the
LHC where one expects mostly SUSY pair production to yield 
sizeable event rates. 

Substantial improvements are expected from the
use of $e^+ e^-$, $\tau^+ \tau^-$ and di-jet plus 
missing momentum event topologies. The results presented
here should encourage one to perform more detailed and
complete background studies and simulations covering other 
R-parity violating signals and improved integrated luminosities
already attained at LEP \cite{new96}. From our point of view
it would be desirable to have additional runs at the Z peak,
considering the fact that large areas of parameter space still 
remain open, where the neutralinos are light enough
to be produced in Z decays.


{\bf Acknowledgements}                      
                                    
We thank Mario Pimenta for useful discussions. This work was 
supported by DGICYT under grant number PB92-0084 and by Acci\'on 
Integrada Hispano-Portuguesa HP-53B, as well as by 
DGICYT (M. A. G. J.) and CNPq (M. B. M. and F. de Campos)
Fellowships.


{\large \bf Appendix}

\appendix

\section{3-Body Neutralino Decays }

In models with spontaneous R-parity violation, the neutral fermions
eigenstates are mixed. We denote them collectively by $\chi^0_i$ with
$(i=1,\ldots,7)$. The neutrinos correspond in this notation to
$i=1,2,3$ and the usual neutralinos to $i=4,\ldots,7$. In the charged
fermion sector, the leptons are mixed with the charginos of the
MSSM. We denote them by $\chi^-_i$ with $(i=1,\ldots,5)$. Again, the
indices $(i=1,2,3)$ correspond, respectively to $e^-$, $\mu^-$ and
$\tau^-$, while the indices $i=4,5$ denote the usual charginos. The
quarks do not mix with the other fermions. We will consider separately
the cases where {\sl all} the particles in the final state have the
same charge and the case where they can have different charges.

\subsection{Same-Charge Final States}

Here the decay we consider is into neutral final states
\begin{equation}
\chi^0_i \ra \chi^0_j +\chi^0_k +\chi^0_m 
\end{equation}
\noi
and proceeds via the neutral current. Due to the Majorana nature of the
neutral fermions, there are 3 distinct diagrams giving rise to some
interference terms. The final result for the decay, within the
approximation that we neglect all masses of the final fermions, is
given by
\begin{eqnarray}
&&\Gamma^{3b}(M_i,O''_L,O''_R) 
=\frac{G^2_F M^5_i}{48\ \pi^3}\
\left[ \vbox to 16 pt{}
 \left( c_1+c_2+c_3 \right) f(x_Z) \hskip 5cm \right.  \cr
&& \vbox to 18 pt{} \left.
\hskip 5 cm \vbox to 16 pt {}
+2 \left( c_{12}+  c_{13}+  c_{23}\right) g(x_Z,x_Z)\right] 
\frac{1}{S_F} 
\end{eqnarray}
\noi
where 
\begin{eqnarray}
c_1&=&\left( O^{'' 2}_{L j i} +O^{'' 2}_{R j i} \right) 
\left( O^{'' 2}_{L k m} +O^{'' 2}_{R k m} \right) \cr
\vbox to 18 pt{}
c_2&=&\left( O^{'' 2}_{L k i} +O^{'' 2}_{R k i} \right) 
\left( O^{'' 2}_{L j m} +O^{'' 2}_{R j m} \right) \cr
\vbox to 18 pt{}
c_3&=&\left( O^{'' 2}_{L k j} +O^{'' 2}_{R k j} \right) 
\left( O^{'' 2}_{L i m} +O^{'' 2}_{R i m} \right) \cr
\vbox to 18 pt{}
c_{12}&=& O''_{L k m} O''_{L j i} O''_{L k i} O''_{L j m}+
 O''_{R k m} O''_{R j i} O''_{R k i} O''_{R j m} \cr
\vbox to 18 pt{}
c_{13}&=&- \left( O''_{L k m} O''_{R j i} O''_{L k j} O''_{L i m}+
 O''_{R k m} O''_{L j i} O''_{R k j} O''_{R i m}\right) \cr
\vbox to 18 pt{}
c_{23}&=& O''_{L k i} O''_{L k j} O''_{R j m} O''_{R i m}+
O''_{R k i} O''_{R k j} O''_{L j m} O''_{L i m} 
\end{eqnarray}
\noi 
$S_F$ is the symmetry factor for identical particles in the final state,
$x_Z= M_i/M_Z$, and the functions $f(x)$ and $g(x,y)$ are given in the
Appendix of ref.~\cite{RPLHC}. The coupling matrices are given in
section 3.

\subsection{Different-Charge Final States}

In this case we consider 3-body final states where two 
are oppositely charged, that is
\begin{equation}
\chi^0_i \ra \chi^0_j +\chi^-_k +\chi^+_m 
\end{equation}
\noi
This decay can proceed via both charged and neutral current. Also in the
charged current case, the Majorana nature of the neutralino implies
the existence of two distinct diagrams. The final result for the
decay, within the approximation that we neglect all masses of the  
final fermions, is given by
\begin{eqnarray}
&&\Gamma^{3b'}(M_i,O''_L,O''_R,O'_L,O'_R,K_L,K_R)\hskip 5cm \cr
&&\vbox to 24 pt {} \hskip 0.5 cm
=\frac{G^2_F M^5_i}{48\ \pi^3}\
\left[ \vbox to 16 pt{}
 c_1  f(x_Z) + \left( c_2+c_3 \right) f(x_W) \right. \cr
&& \vbox to 18 pt {}
\left. \vbox to 16pt {}
\hskip 3 cm
+2 \left( c_{12}+  c_{13}\right) g(x_Z,x_W) 
+  2 c_{23} g(x_W,x_W) \right] 
\label{diffcharge}
\end{eqnarray}
\noi
where $x_W=M_i/M_W$, the functions  $f(x)$ and $g(x,y)$ are 
as before, and the coefficients $c_i$ are now given by
\begin{eqnarray}
c_1&=&\left( O^{'' 2}_{L j i} +O^{'' 2}_{R j i} \right) 
\left( O^{' 2}_{L k m} +O^{' 2}_{R k m} \right) \cr
\vbox to 18 pt{}
c_2&=&\left( Y^2_{L k i} +Y^2_{R k i} \right) 
\left( X^2_{L j m} +X^2_{R j m} \right) \cr
\vbox to 18 pt{}
c_3&=&\left( Y^2_{L k j} +Y^2_{R k j} \right) 
\left( X^2_{L i m} +X^2_{R i m} \right) \cr
\vbox to 18 pt{}
c_{12}&=& O'_{L k m} O''_{L j i} Y_{L k i} X_{L j m}+
 O'_{R k m} O''_{R j i} Y_{R k i} X_{R j m} \cr
\vbox to 18 pt{}
c_{13}&=&- \left( O'_{L k m} O''_{R j i} Y_{L k j} X_{L i m}+
 O'_{R k m} O''_{L j i} Y_{R k j} X_{R i m}\right) \cr
\vbox to 18 pt{}
c_{23}&=& Y_{L k i} Y_{L k j} X_{R j m} X_{R i m}+
Y_{R k i} Y_{R k j} X_{L j m} X_{L i m}
\end{eqnarray}
\noi
with
\begin{eqnarray}
Y_{L i j}&=&\frac{1}{\sqrt{2}}\  K_{L i j} \cr
\vbox to 16 pt {}
Y_{R i j}&=&\frac{1}{\sqrt{2}}\  K_{R i j} \cr
\vbox to 16 pt {}
X_{L i j}&=&\frac{1}{\sqrt{2}}\  K_{L j i} \cr
\vbox to 16 pt {}
X_{R i j}&=&\frac{1}{\sqrt{2}}\  K_{R j i} 
\end{eqnarray}
\noi
In all the previous expressions, the coupling matrices are taken to be
real. The sign factors $\epsilon_i$ and $\eta_i$ are introduced as
explained in section 3.

\noi
Before we close this discussion, we notice that Eq.~\ref{diffcharge}
can also be applied, with obvious modifications,  to the case of
quarks in the final state. For definiteness consider the process
\begin{equation}
\chi^0_i \ra \chi^0_j + u + \overline{u}
\end{equation}
\noi
It is clear that in this case we have only the $Z$ exchange
diagram. The formula of Eq.~\ref{diffcharge} it is still  valid
with the following values for the coefficients $c_i$
\begin{eqnarray}
c_1&=&\left( O^{'' 2}_{L j i} +O^{'' 2}_{R j i} \right) 
\left[ \left(\frac{1}{2} -Q_u \sin^2 \theta_w \right)^2  +
\left( -Q_u \sin^2 \theta_w \right)^2 
\right] \cr
\vbox to 18 pt{}
c_2&=&c_3=c_{12}=c_{13}=c_{23}=0 
\end{eqnarray}
Notice that the neutralino decay formulas given above correct those
previously given in the Appendix of ref.~\cite{RPLHC}, which were not 
correct for the case of decays involving Majorana fermions, such as 
$\chi \ra 3 \nu_\tau$. However the functions  $f(x)$ and $g(x,y)$ are 
the same as in ref.~\cite{RPLHC}.

\newpage
\begin{table}
\label{decays}
\begin{center}
\begin{tabular}{||l|l||}\hline\hline
$\chi\chi$ & $\chi\nu$ \\ \hline
 $p\!\!\!/_T+l_i^+l_i^- +l_j+l_j-$ & $p\!\!\!/_T + l_i^+l_i^-$  \\
 $p\!\!\!/_T+l_i^+l_i^- + 2 jets$ & $p\!\!\!/_T + (\tau \ l)$ \\
 $p\!\!\!/_T+l_i^+l_i^- +(2 jets + \tau)$ & $p\!\!\!/_T+\tau+ 2jets$ \\
 $p\!\!\!/_T+l_i^+l_i^-+(\tau l)$ & $p\!\!\!/_T+2jets$ \\
 $p\!\!\!/_T+l_i^+l_i^-$ & \\
 $p\!\!\!/_T + 4 jets$ & \\
 $p\!\!\!/_T + 4 jets+\tau$ & \\
 $p\!\!\!/_T + 2 jets+(\tau l)$ & \\
 $p\!\!\!/_T + 2 jets$ & \\
 $4 jets +\tau \tau$ & \\
 $p\!\!\!/_T +2 jets +\tau$ & \\
 $p\!\!\!/_T +2 jets +\tau (\tau \ l)$ & \\
 $p\!\!\!/_T (\tau \ l)(\tau \ l)$ & \\
 $p\!\!\!/_T (\tau \ l)$ & \\ \hline\hline
\end{tabular}
\end{center}
\caption{Final signals arising from neutralino 
pair-production (left column) as well single production (right column).
The total charge of the particles between parenthesis must be zero.}
\end{table}

\newpage

\bef
\centerline{\protect\hbox{
\psfig{file=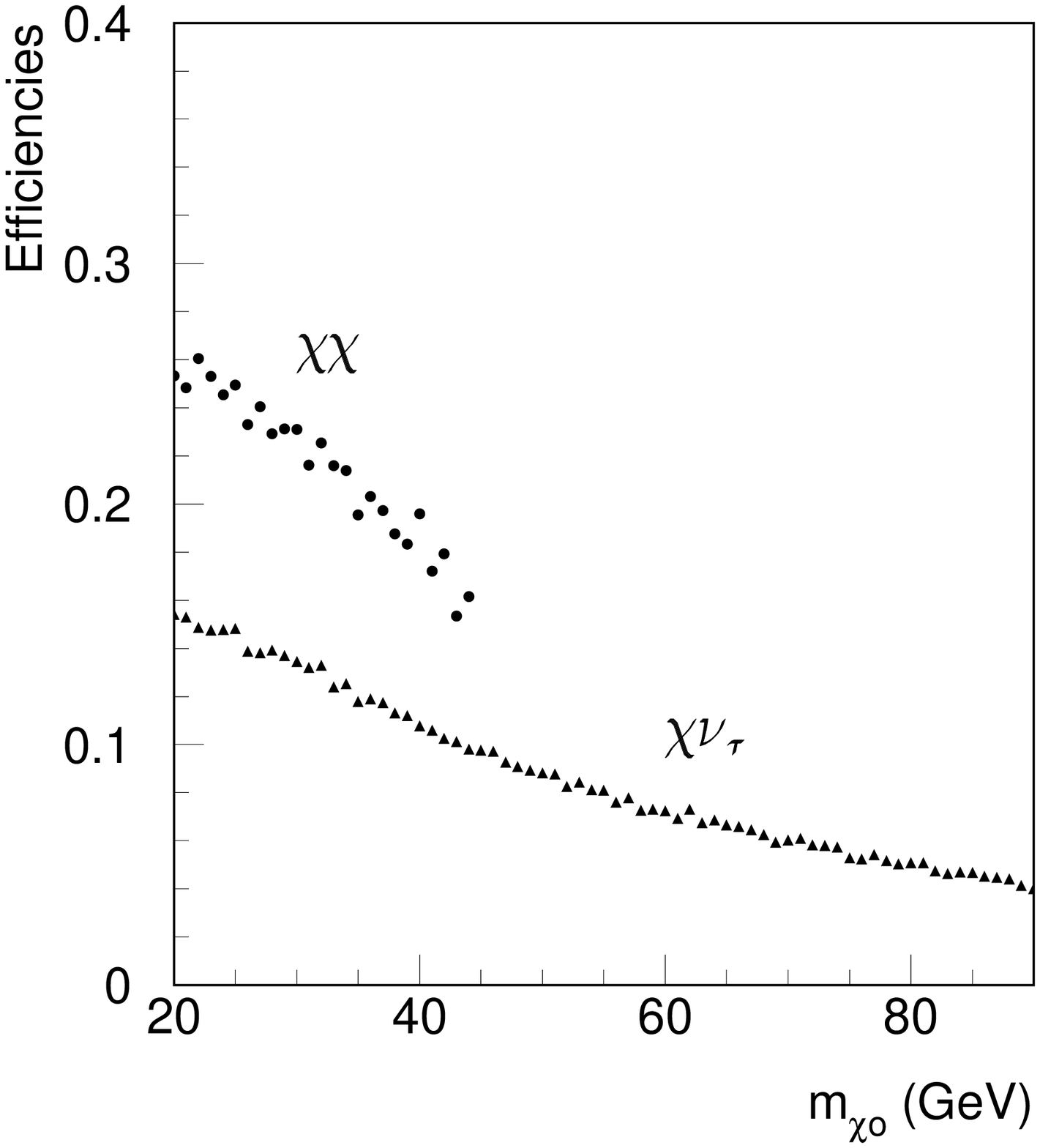,height=10cm}}}
\caption{Detection efficiencies for the \ptmm signal associated to 
$\chi\nu$ and $\chi\chi$ production channels.}
\label{0}
\eef

\bef
\centerline{\protect\hbox{
\psfig{file=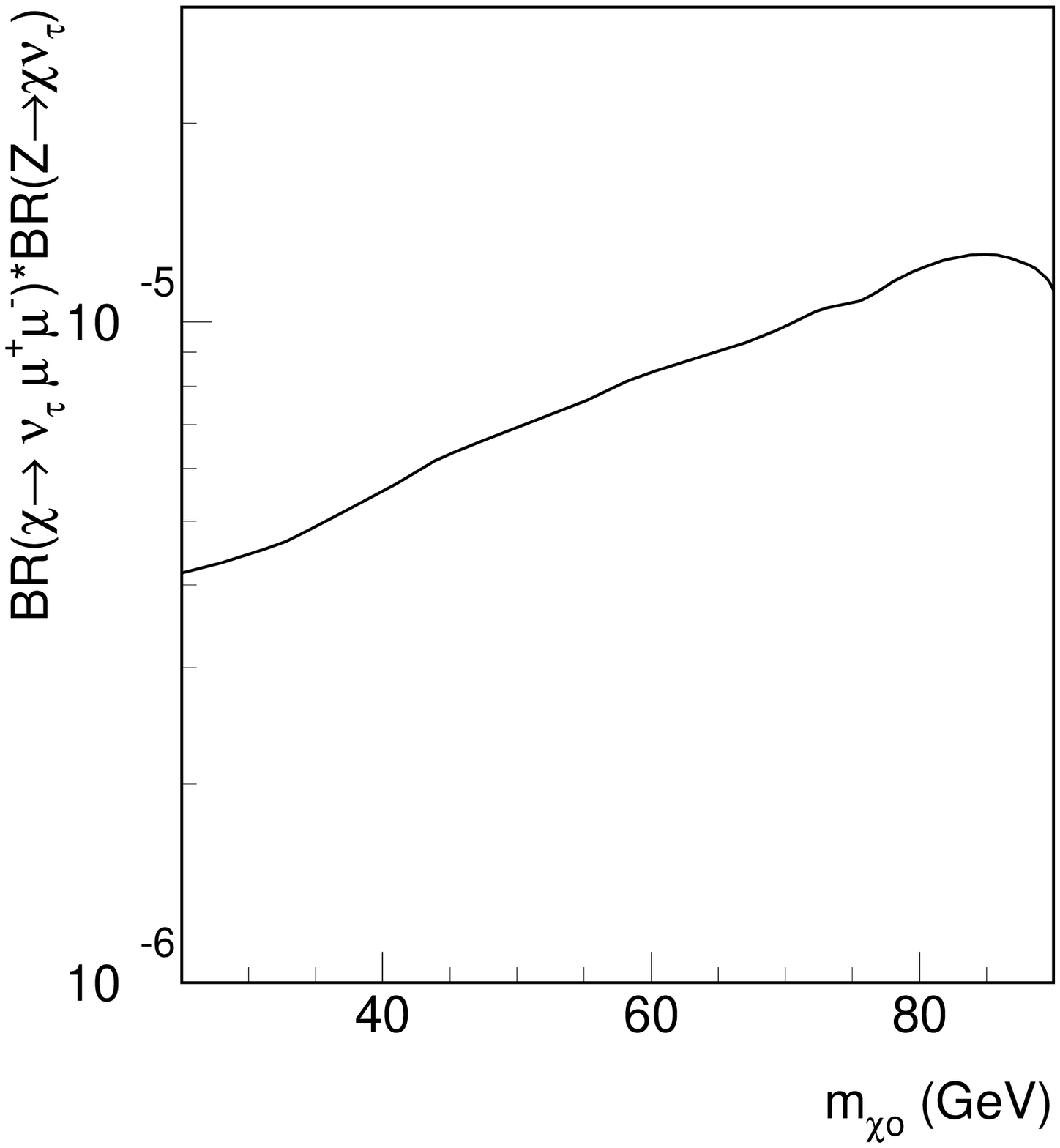,height=10cm}}}
\caption{Region of sensitivity obtained at the $95 \%$ C.L. for 
$BR (Z \ra \chi \nu) BR(\chi \rightarrow \mu^+\mu^- \nu)$, as a 
function of the lightest neutralino mass $m_{\chi^0}$. This is 
derived from searches of $ p\!\!\!/_T +\mu^+ \mu^- $ events that 
would arise from single neutralino production at LEP, followed by 
$\chi \ra \mu^+ \mu^- \nu$ decays.}
\label{I}
\eef

\bef
\centerline{\protect\hbox{
\psfig{file=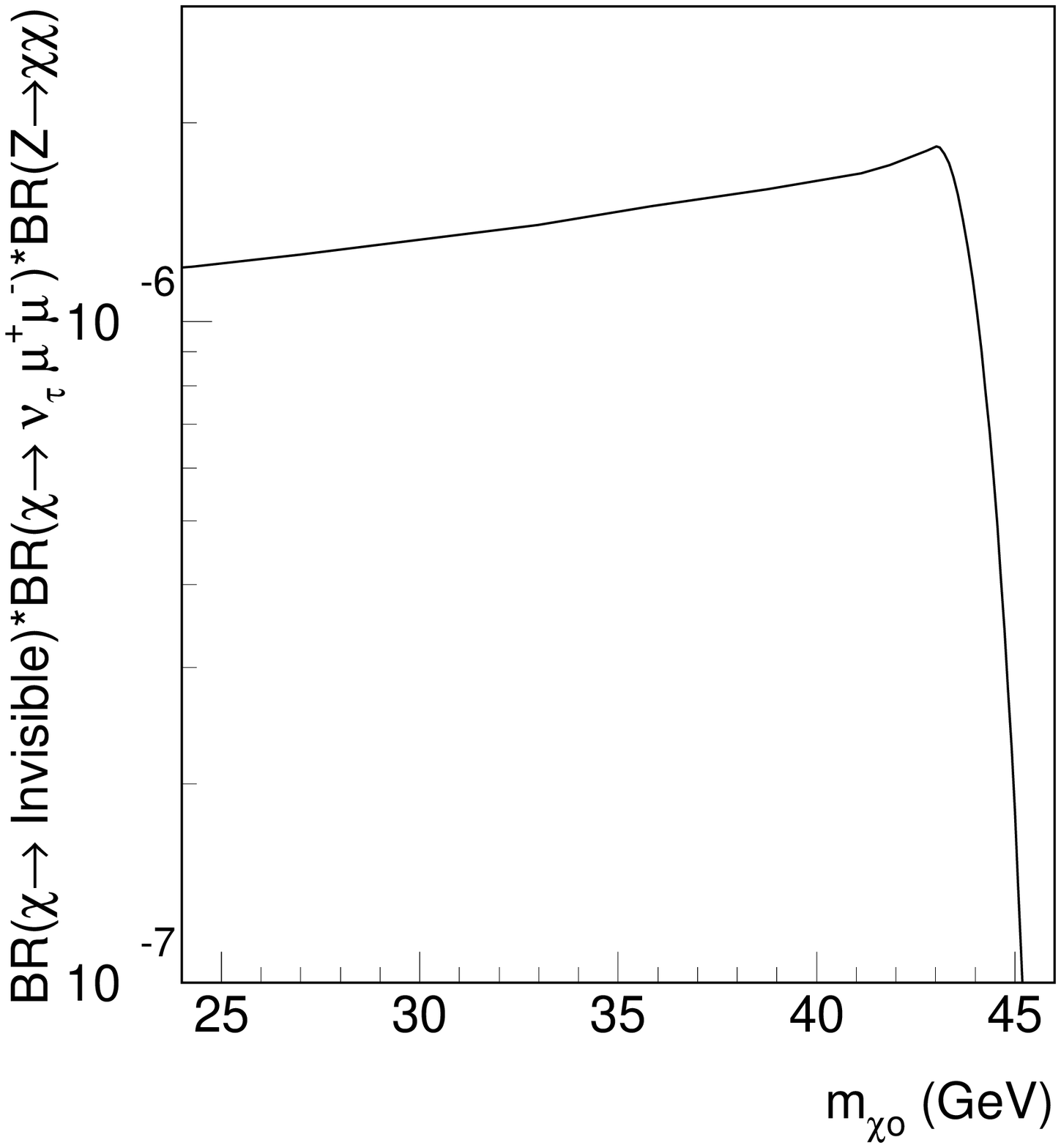,height=10cm}}}
\caption{Region of sensitivity obtained at the $95 \%$ C.L. for 
$BR (Z \ra \chi \chi) (\chi \rightarrow \mu^+\mu^- \nu)
BR(\chi \ra \mbox{invisible})$ as a function of the lightest 
neutralino mass. This is derived from searches of 
$ p\!\!\!/_T +\mu^+ \mu^- $ events that would
arise from neutralino pair production at LEP, with one neutralino
decaying invisibly and the other decaying as $\chi \ra \mu^+ \mu^- \nu$.}
\label{II}
\eef

\bef
\centerline{\protect\hbox{
\psfig{file=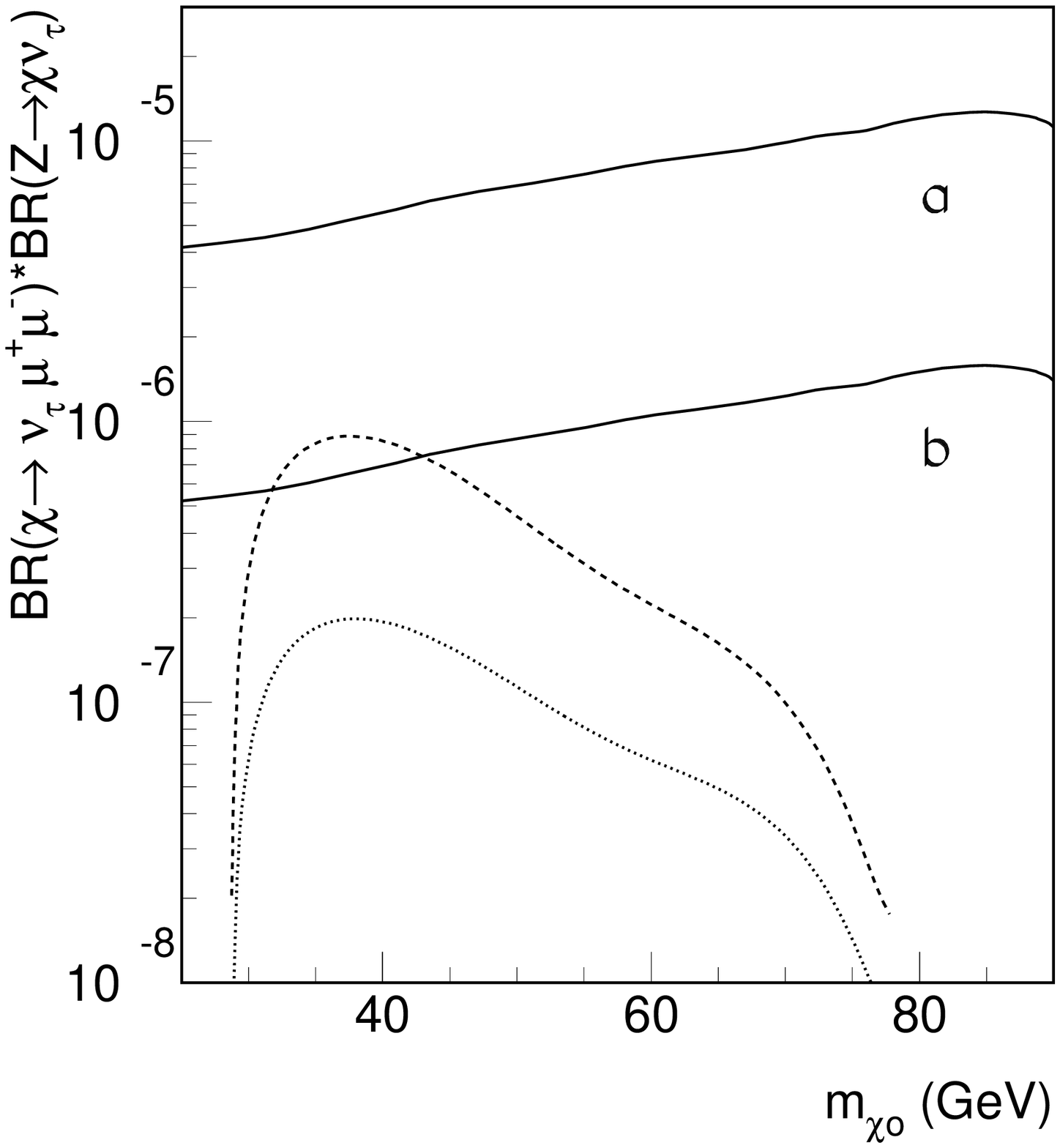,height=10cm}}}
\caption{Comparison of the attainable limits on
$BR (Z \ra \chi \nu) BR(\chi \rightarrow \mu^+\mu^- \nu)$
versus the lightest neutralino mass, with the maximum theoretical values 
expected in different R-parity breaking models. The solid line (a) is the 
same as in Fig. 1, while (b) corresponds to the improvement expected
from including the $e^+e^-\nu$ channel, as well as the 
combined statistics of the four LEP experiments.
The dashed line corresponds to the model of ref. [16]
allowing \mnt values as large as the present laboratory limit of
ref. [23], while the dotted one is calculated in the 
spontaneous R-parity-violation model of ref. [5].}
\label{III}
\eef

\bef
\centerline{\protect\hbox{
\psfig{file=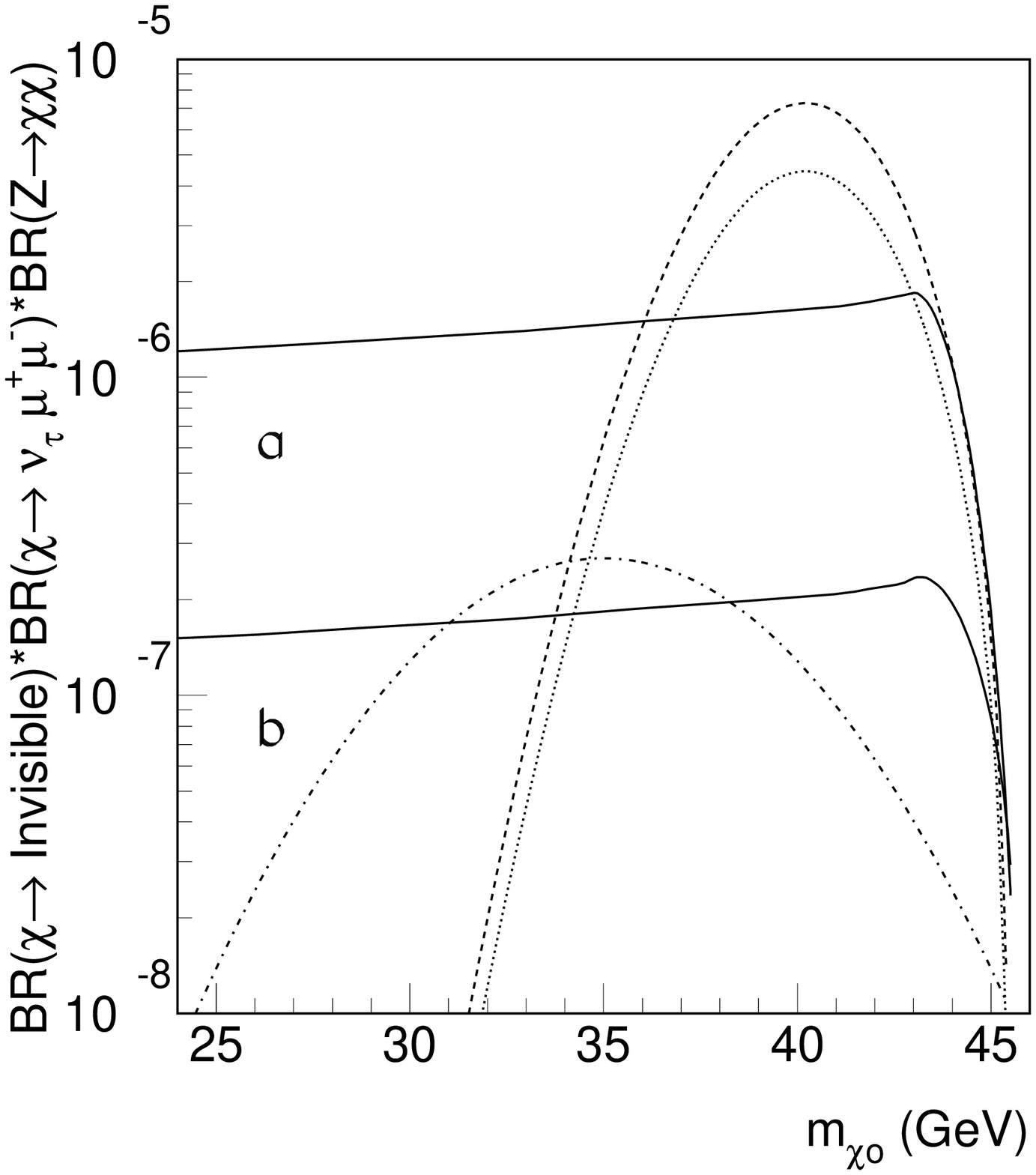,height=10cm}}}
\caption{Comparison of the attainable limits on
$BR (Z \ra \chi \chi) BR(\chi \rightarrow \mu^+\mu^- \nu)$
versus the lightest neutralino mass, with the 
maximum theoretical values expected in different 
R-parity breaking models. The solid line (a) is the same as in
Fig. 2, while (b) corresponds to the improvement expected
from including the $e^+e^-\nu$ channel, as well as the 
combined statistics of the four LEP experiments.
The dashed line corresponds to the model in ref. [16]
allowing \mnt values as large as the present limit,
the dotted one does implement the restriction on \mnt
suggested by nucleosynthesis, and the dash-dotted one is 
calculated in the model of ref. [5].}
\label{IV}
\eef

\bef
\centerline{\protect\hbox{
\psfig{file=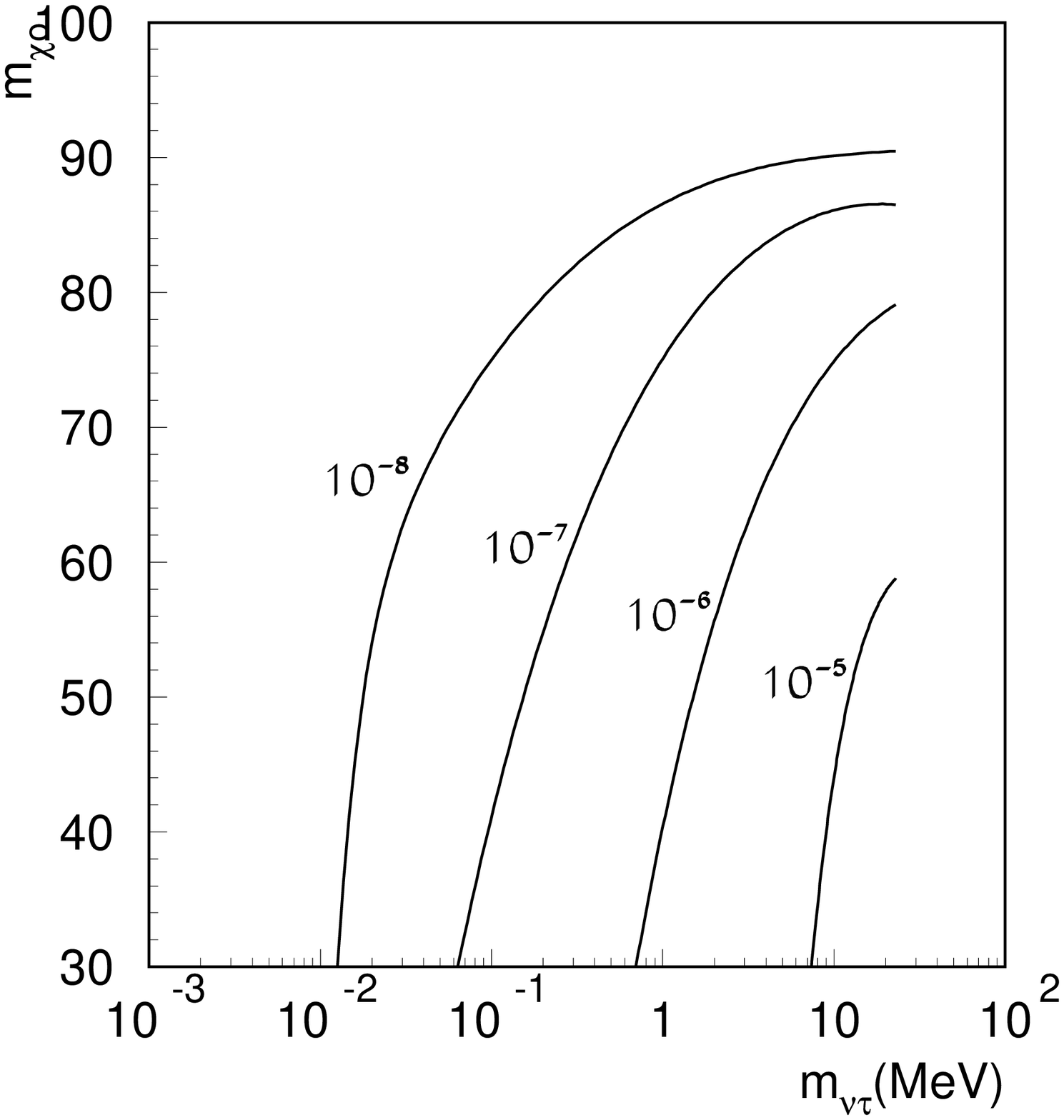,height=10cm}}}
\caption{Attainable values for the R-parity breaking strength
$BR (Z \rightarrow \chi \nu )$ versus the $\tau$-neutrino mass
in the model of ref. [5].}
\label{V}
\eef

\end{document}